# Erwin L. Hahn

1921-2016

**A Biographical Memoir by Alexander Pines and Dmitry Budker**

Erwin Louis Hahn was one of the most innovative and influential physical scientists in recent history, impacting generations of scientists through his work in nuclear magnetic resonance (NMR), optics, and the intersection of these two fields. Starting with his discovery of the spin echo, a phenomenon of monumental significance and practical importance, Hahn launched a major revolution in how we think about spin physics, with numerous implications to follow in many other areas of science. Students of NMR and coherent optics quickly discover that many of the key concepts and techniques in these fields derive directly from his work.

## Biography in brief

Erwin Hahn was born on June 9, 1921 in Farrell (near Sharon), Pennsylvania. His mother Mary Weiss, daughter of a rabbi in Vrbas (originally in Austria-Hungary, now in Serbia), was sent by her family to the United States to marry the son of family acquaintances, Israel Hahn, whom she had never met. The marriage produced six boys and one girl (Deborah, who died in the 1918 flu epidemic); Erwin was the youngest. The fact that he had five brothers certainly influenced his development. The family home was in Sewickley, Pennsylvania, and Israel Hahn operated several dry cleaning stores in the area. Neither the business nor the marriage proved to be a success: Israel's presence at home was erratic; he made bad business decisions; and he had phases of authoritarian religious fanaticism which created tensions. Eventually, Israel abandoned the family, leaving Mary to parent Erwin and his brothers as well as run the one remaining dry cleaning shop from the business. His oldest brother Simon went to work for a relative to help support the family. It was Simon who bought his little brother a chemistry set; Hahn recalled distilling, then spilling, urine on the windowsill to the despair of his mother. Economic necessity meant that only Erwin and his slightly older brother Philip were able to go to college, both on scholarships (Philip became a professor of economics).

Hahn received his BSc in Chemistry in 1943 from Juniata College and afterwards, impressed by the fundamental concepts of physics, completed a year of graduate studies in Physics at Purdue University. Hahn's studies were interrupted by war-time service in the US Navy, where he was a radar and sonar instructor. He continued his studies in Physics at the University of Illinois, earning his MSc in 1947 and his DSc in 1949.

Hahn had several possibilities for a career, among them the Navy, music (he was an enthusiastic and gifted violinist), as well as acting and comedy. In the Navy he was frequently on stage as an entertainer and was asked if he would go professional. His older brother Milton was also a professional actor for some years. However, Hahn turned towards science, in particular magnetic resonance and, later, optics.

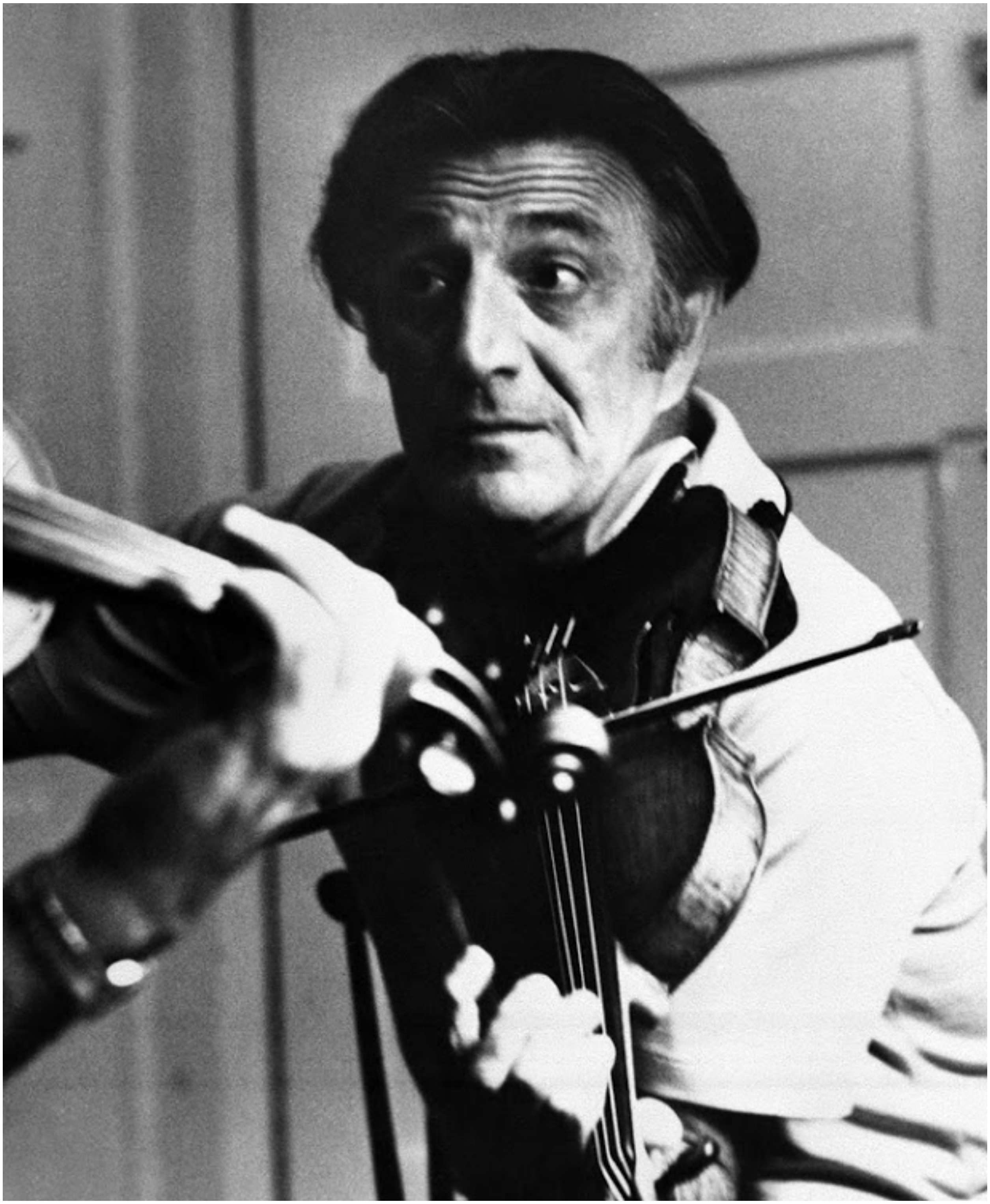

*Figure 1. Erwin L. Hahn in the early 1970s.*

He was a research associate at the University of Illinois; a National Research Council fellow at Stanford University, where he worked with Felix Bloch; and a research physicist at IBM Watson Scientific Computing Laboratory. In 1955 he came to the University of California, Berkeley, where he joined the physics faculty, and where he remained until his death on September 20, 2016 at the age of 95.

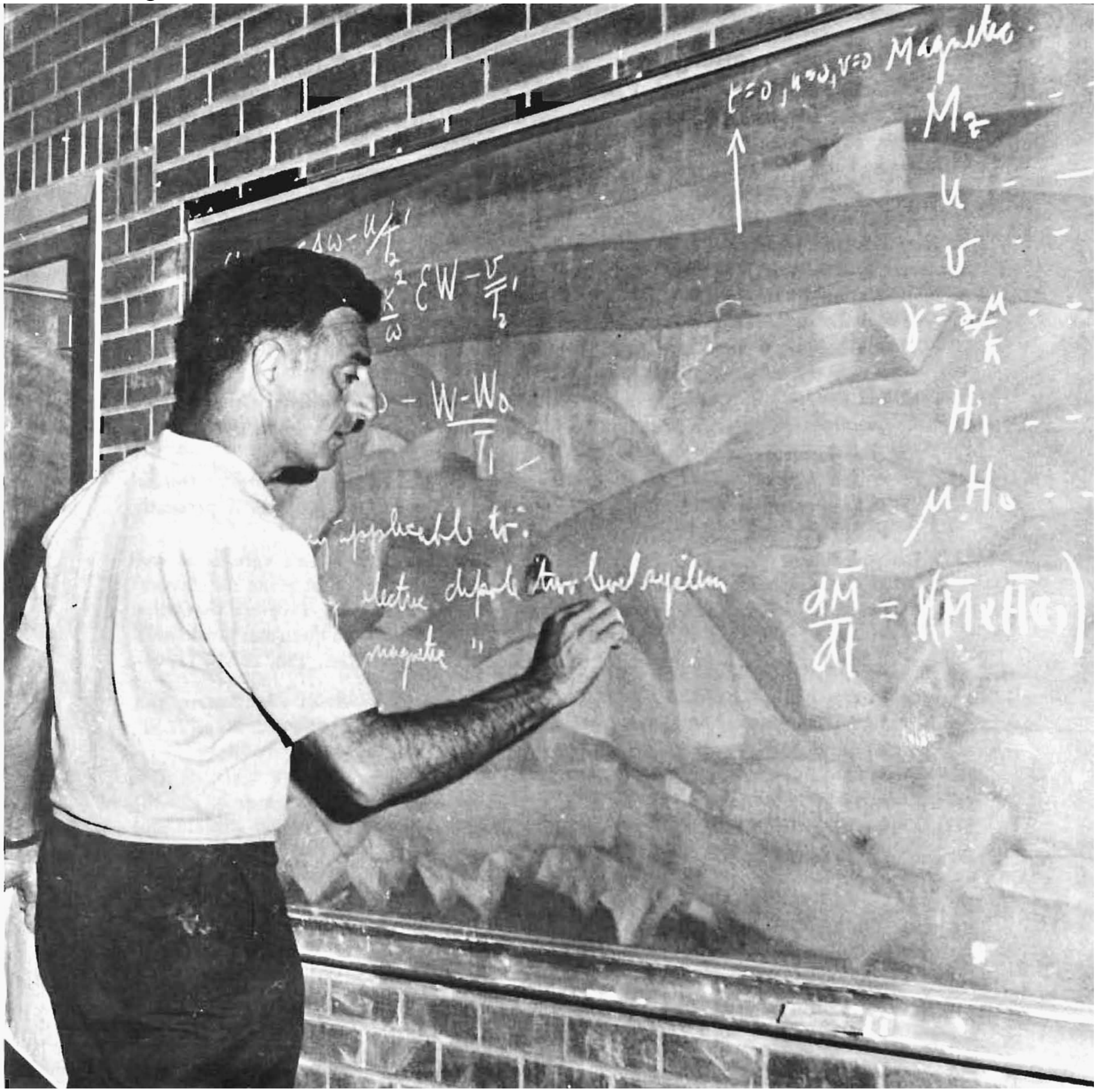

*Figure 2. Hahn at work (1960s).*

Hahn was married twice, to Marian Ethel Failing in 1944 and, after her death in 1978, to Natalie Woodford Hodgson in 1980. The contributions of both Marian and Natalie to Hahn's career were significant and profound. He was proud to have a son in medicine, and of both of his daughters' professions and inclination for music. His colleagues have noted that Hahn himself was deeply into music, and if not for physics, music would have likely been his career. Hahn is survived by his wife Natalie, three children, two stepchildren, three grandchildren and four great-grandchildren.

# Major scientific contributions

Some of the most significant contributions of E. L. Hahn, who was referred to as "the Wizard of Magnetic Resonance" include:

- The introduction of nuclear free-induction decay (FID) which was reported in a short Physical Review paper that was published in 1950 just slightly ahead of Hahn's more famous paper on spin echoes. In FID, polarized nuclei are subject to a pulse of a driving magnetic field but are then allowed to evolve "freely," while their magnetization is monitored. While Hahn downplayed the significance of this work as "obvious," it has in fact become the basis of pulsed NMR, the dominant modern approach in NMR.
- Hahn was happy to tell the story of his "accidental" discovery of the spin echo, the most famous of his achievements. He was a post-doctoral fellow when he discovered spin echoes, though he emphasized that at that time he was given freedom to work more like an independent research scientist. While studying nuclear spin coherence relaxation, he applied not one but two driving pulses separated by a time interval. What he saw (and what he at first thought to be a "glitch") was that, in addition to decaying FID signals following each of the two separate pulses, there was a third "ghost" signal appearing at a time following the second pulse equal to the time separation between the two pulses. This was puzzling because it was not clear where the signal came from; the echo occurred long after the FID following each of the pulses died out due to dephasing, then thought of as an irreversible process in the thermodynamic sense. Eventually, Hahn recognized that the "echo" signal was due to "refocusing" of the precession of different nuclei in the sample occurring at slightly different frequencies due to magnetic-field inhomogeneities and that the system's order was "hidden" but not gone. The second pulse, in effect, creates a kind of time reversal, where the relative phases accumulated by the spins during the evolution between the pulses are undone during the evolution after the second pulse. Today, the spin echo and its countless generalizations, for instance to sequences of not two but up to thousands of pulses, constitute the basis of essentially all magnetic resonance applications, including the familiar medical-diagnostic magnetic resonance imaging (MRI).
- Indirect dipole-dipole (scalar or *J*-) coupling between nuclear spins in a single molecule was independently discovered by two teams: Hahn and D. E. Maxwell, and H. S. Gutowsky, D. W. McCall, and C. P. Slichter. While the usual interaction between two magnetic dipoles (magnetic nuclei) averages to zero when the molecule in a liquid or a gas rapidly tumbles, altering the relative positions of the nuclei, the two teams discovered that a part of the dipole-dipole interaction survives the tumbling. In modern language, the origin of the effect is second-order hyperfine interaction involving molecular electrons. These scalar couplings went on to become part of the "molecular fingerprint" widely used in NMR spectroscopy. Moreover, the growing subfield of zero- and ultralow-field (ZULF) NMR spectroscopy (in which both authors of this memoir currently work and to which Hahn was one of the early contributors) is entirely based on indirect couplings.
- While Hahn did not discover nuclear quadrupole resonance (NQR), a kind of magnetic resonance that occurs for nuclei with spin >1/2 that comes about due to the interaction of these nuclei with electric-field gradients in solids, he literally "wrote the book" on the subject. Recent applications of NQR include, apart from chemical analysis, detection of toxic and explosive substances. An important property of NQR is that it does not require application of a strong magnetic field. Later work with A. Pines, J. Clarke, and A.

Trabesinger led to the development of a more general ZULF NMR technique that similarly does not require any strong fields and is colloquially known as "NMR without magnets."

- Together with D. E. Kaplan, Hahn introduced spin-echo double resonance (SEDOR) techniques, which are widely used to study interactions between dissimilar nuclei and, for instance, to measure the lengths of chemical bonds. SEDOR was one of the first "2D NMR" methods in that it studied the spin-system evolution as a function of two different time-delay intervals in an experimental sequence. SEDOR was a precursor of the powerful multi-dimensional NMR methods used in protein-structure analysis today.
- Hahn pioneered the study of nuclear spin noise and elucidated the physics of radiation damping in a spin system coupled to a resonator or a microwave cavity.
- Together with S. R. Hartmann, Hahn devised an ingenious technique involving matching of radio-frequency fields to couple an abundant and a rare nuclear species allowing detection of the rare spins. It later evolved into development of methods for cross-polarization: the transfer of nuclear order from one kind of nuclei to another, greatly expanding the analytical capabilities of NMR. The Hartmann-Hahn results showed that the energy-conservation concept can be extended to the "rotating frame."
- Hahn and S. L. McCall's major discovery in the field of optics was the phenomenon of self-induced transparency whereby a pulse of light propagates without loss in an absorptive medium (which is due to reversible exchange of energy between the light and the medium). This is closely related to the physics of solitary waves (solitons), which were first observed and studied in optics by Hahn and coworkers.
- Together with R. G. Brewer, Hahn introduced a complete theory of coherent two-photon optical processes, a precursor to an active field of research in subsequent decades. This is another beautiful example of how he transferred ideas about coherent transient phenomena in NMR to the field of optics.
- As a pioneer in both fields, Hahn wrote and spoke extensively on the deep (and unfortunately often underappreciated) connection between NMR and quantum optics, offering useful insights to both disciplines.

Beyond his famous published works, Hahn was always a brilliant, provocative, and entertaining lecturer, raconteur and teacher.

## Scientific offspring and influence

Hahn had 51 or so doctoral students. Lawrence Wald, one of the last of these, recalls that Erwin came by the lab every day "like clockwork" and always had a lot of ideas that he was eager to discuss. At times this was not particularly appreciated by the graduate students who wanted to "get things done" instead. While Hahn's ideas were typically original and interesting, nine out of ten of them would eventually be found subtly flawed, and he always needed a "sounding board" to poke holes in them. Hahn was particularly fond of students who were good at poking holes and who would argue scientific points. One of these students (and later, a frequent visitor to Hahn's lab) was Tycho Sleator, whom the Hahn group secretary called "Erwin's illegitimate son" due to the similarity of their personalities. Lawrence Wald recalls that Hahn and Sleator would argue harder than he (Wald) found comfortable.

Most of Erwin's students went to work for Bell Labs after graduation, and in this way his scientific influence proliferated significantly.

Neither of the authors of this memoir are Hahn's direct scientific offspring, but both were greatly influenced by him at different times. Dmitry Budker first met Erwin Hahn almost a quarter-century ago in Novosibirsk (Russia), and was a colleague of his in the UC Berkeley Department of Physics. Alexander Pines---in the Department of Chemistry.

## Hahn's views on science

Hahn emphasized the importance of his technical work in the Navy, where he learned to pulse oscillators based on vacuum-tube technology so that they could be turned on and off (gated) "cleanly." He claimed that this practical knowledge launched his research career.

Throughout his career, Hahn had a deep suspicion of "black-box" devices and preferred to build as much as possible of the experimental apparatus himself. He used to tell how, as a post-doc with Felix Bloch, they built their own oscilloscope. Nobody builds their own oscilloscopes these days, but even the most recent of Hahn's students would recall Hahn grumbling at seeing them play with a newly acquired modern digital oscilloscope: "How do you *really* know what the thing is doing to your signal?"

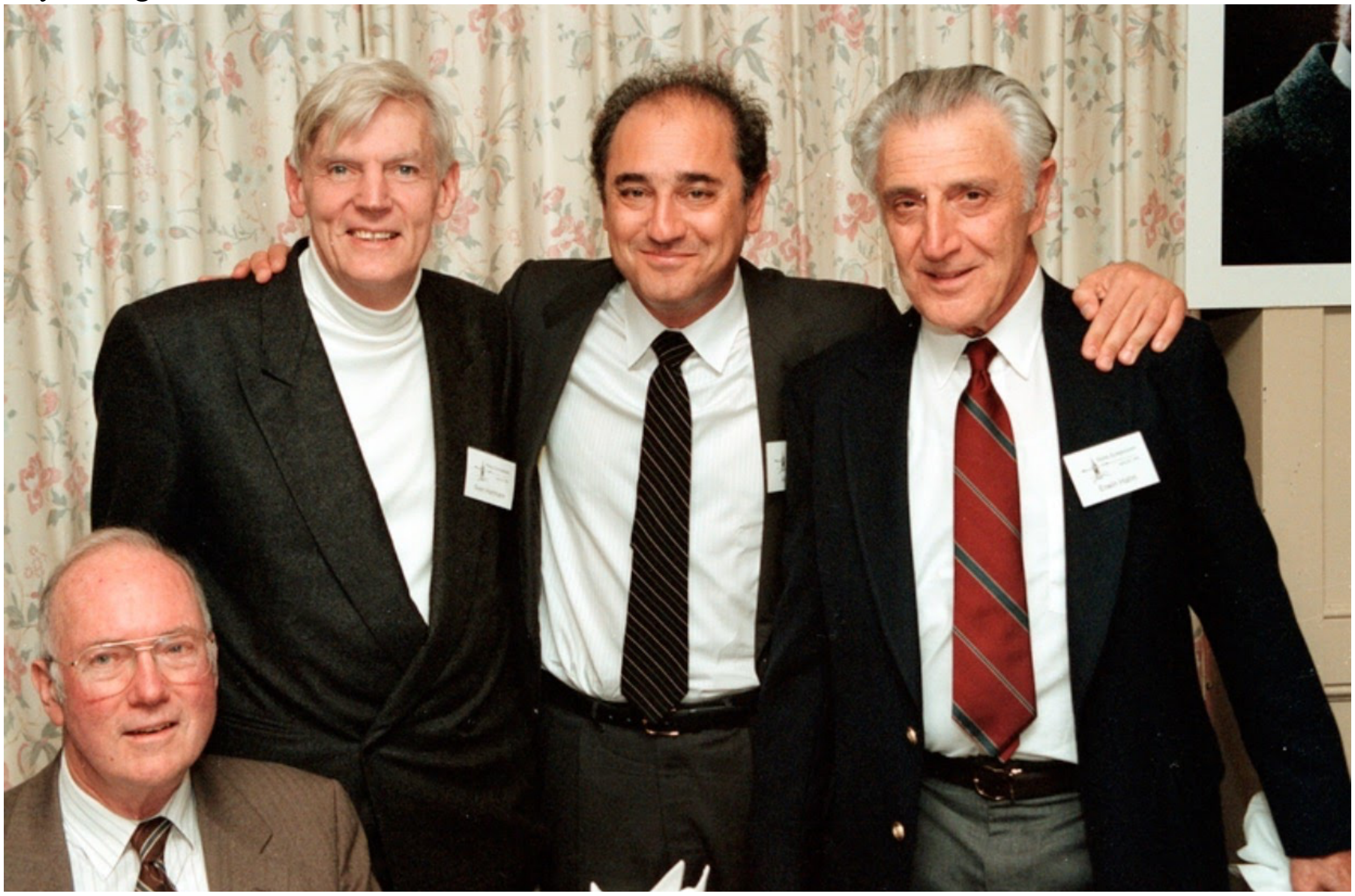

*Figure 3. Colleagues and friends. Left to right: C. H. Townes, S. Hartmann, A. Pines, and E. L. Hahn at a symposium in honor of Hahn's 70th birthday* organized by Berkeley faculty members A. Pines (Chemistry), Walter Knight (Physics) and Melvin V. Klein (Biology).

## Recognition

E. L. Hahn's contributions to science and technology are widely recognized. A partial list of his awards and honors includes:

- Guggenheim Fellowships (1961 and 1969)
- Buckley Prize in Solid State Physics, American Physical Society (1971)
- Prize of the International Society of Magnetic Resonance (1971)
- Alexander von Humboldt Award (1976)
- Wolf Foundation Prize in Physics (1983)
- California Inventors Hall of Fame (1984)
- US Department of Energy Award for Sustained Research in DC Squid NMR (1986)
- Berkeley Citation, University of California at Berkeley (1991)
- US National Academy of Sciences Comstock Prize for discoveries in electricity, magnetism or radiation (1993)
- Honorary Doctor of Science, University of Stuttgart , Germany (2001)
- Russell Varian Prize (2004)
- Honorary Doctorate, Oxford University (2009)
- Gold Medal of the International Society of Magnetic Resonance in Medicine (2016).

His honorary memberships included:

- Fellow, American Physical Society (1952)
- Fellow, American Academy of Arts and Sciences (1971)
- U.S. National Academy of Sciences (1972)
- Foreign Member, Slovenian Academy of Sciences (1981)
- Honorary Fellow, Brasenose College, Oxford (1982)
- Foreign Member, French Academy of Sciences (1992)
- Honorary Member, International Society of Magnetic Resonance (1995)
- Honorary Fellow, International Society of Paramagnetic Resonance (1996)
- Berkeley Society of Fellows (1999)
- Honorary Fellow, International Society of Magnetic Resonance in Medicine (2000)
- Foreign Member, Royal Society, U.K. (2000)
- Member, Institute of Physics, U.K. (2000).

Hahn did not receive a Nobel Prize, despite the huge impact his work had across many disciplines benefitting humanity, and in particular, analytical spectroscopy, quantum spin physics, and modern biological imaging.

Although he never worked directly on MRI methodology, the impact of his foundational work in magnetic resonance is well appreciated. It is memorialized in the Erwin L. Hahn Institute for Magnetic Resonance Imaging, founded in July 2005 by the University Duisburg-Essen (Germany) and the Radboud University Nijmegen (The Netherlands). This institute opened in 2006 in Essen, Germany, and boasts a 7 tesla whole-body magnetic-resonance imager (MRI). In contrast, typical medical MRI operate at a magnetic field strength of 1.5 tesla. Erwin could not attend the opening of the Hahn Institute due to illness, and asked his former doctoral student Lawrence Wald to represent him.

## Hahn stories

Hahn had a remarkable and legendary sense of humor which drew on all levels of decorum, and he was always glad to share “Hahn stories, anecdotes, and witty remarks” with colleagues and friends. While they were much appreciated by persons close to him, his wit could bother others. In Hahn's case, the sense of humor was accompanied by sensitivity and (at the very least, honestly attempted) tact.

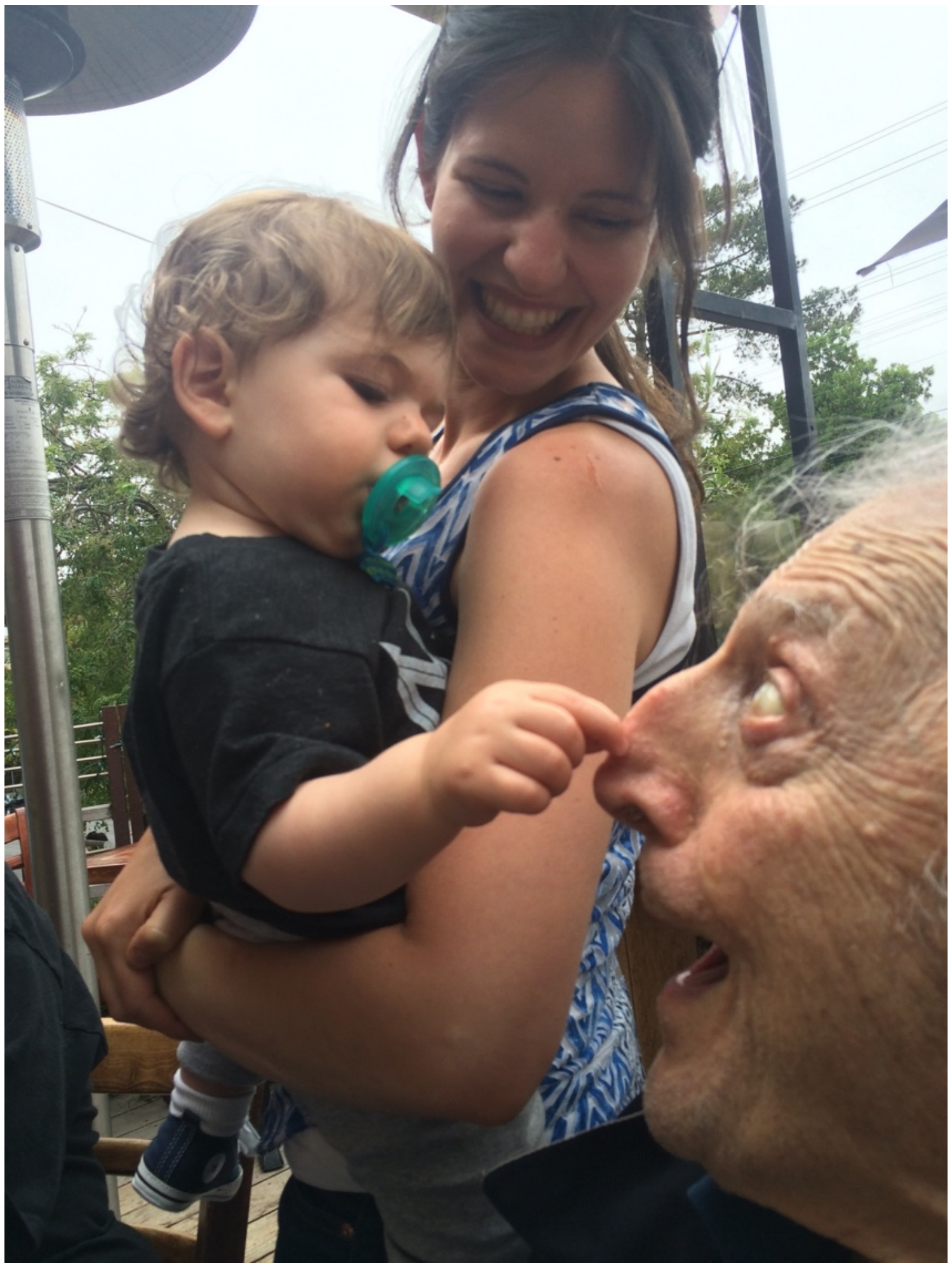

*Figure 4 Erwin with granddaughter-in-law Ashley Hahn and great grandson Hudson Hahn (circa 2015).*

We recall several anecdotes to give the reader a feel for Hahn's fast, biting wit.

Some years ago, Hahn was walking on the Berkeley campus with Alexander Pines, whose nose was bandaged and bloody because of an injury. They encountered a student of Alex's, who said "What happened to your nose?" Pines began to explain but was cut short by Hahn, who said: "Alex, why do you think he is asking about yours?"

The authors of this memoir have many personal recollections of interactions with Hahn. Here we recall several of these.

Pines remembers: My own first indirect association with Hahn involved an experiment that Won-Kyu Rhim and I did at MIT in the laboratory of the great John S. Waugh, concerning the free induction decay of coupled spins in a crystal. The question then was whether the decay of magnetization under this complex Hamiltonian was really irreversible—it turned out that you could, by applying an extended sequence of pulses called a "magic sandwich" after the total decay of order, bring back the magnetization. About that time, Waugh was going out to California and he planned to drop in on the father of the spin echo and tell him about our work. When he came back to the lab from the trip and we asked him what Hahn had to say, he replied that Hahn had remarked, "With that many pulses I could bring back the Messiah!"

Budker remembers: On several occasions Hahn and I were asked to present a Physics of Music "show-and-tell" lecture at the Physics Department Open House. On one occasion we filled the stage of the large lecture hall with all kinds of acoustic apparatus and unusual instruments. At the beginning of the lecture I introduced Erwin to the audience as a famous professor who would now introduce them to the subject. Erwin came on stage and said:
--"Good morning! I feel a bit like a mosquito at a beach…", and indicated the numerous props to the startled audience.

--"I don't know where to start!"

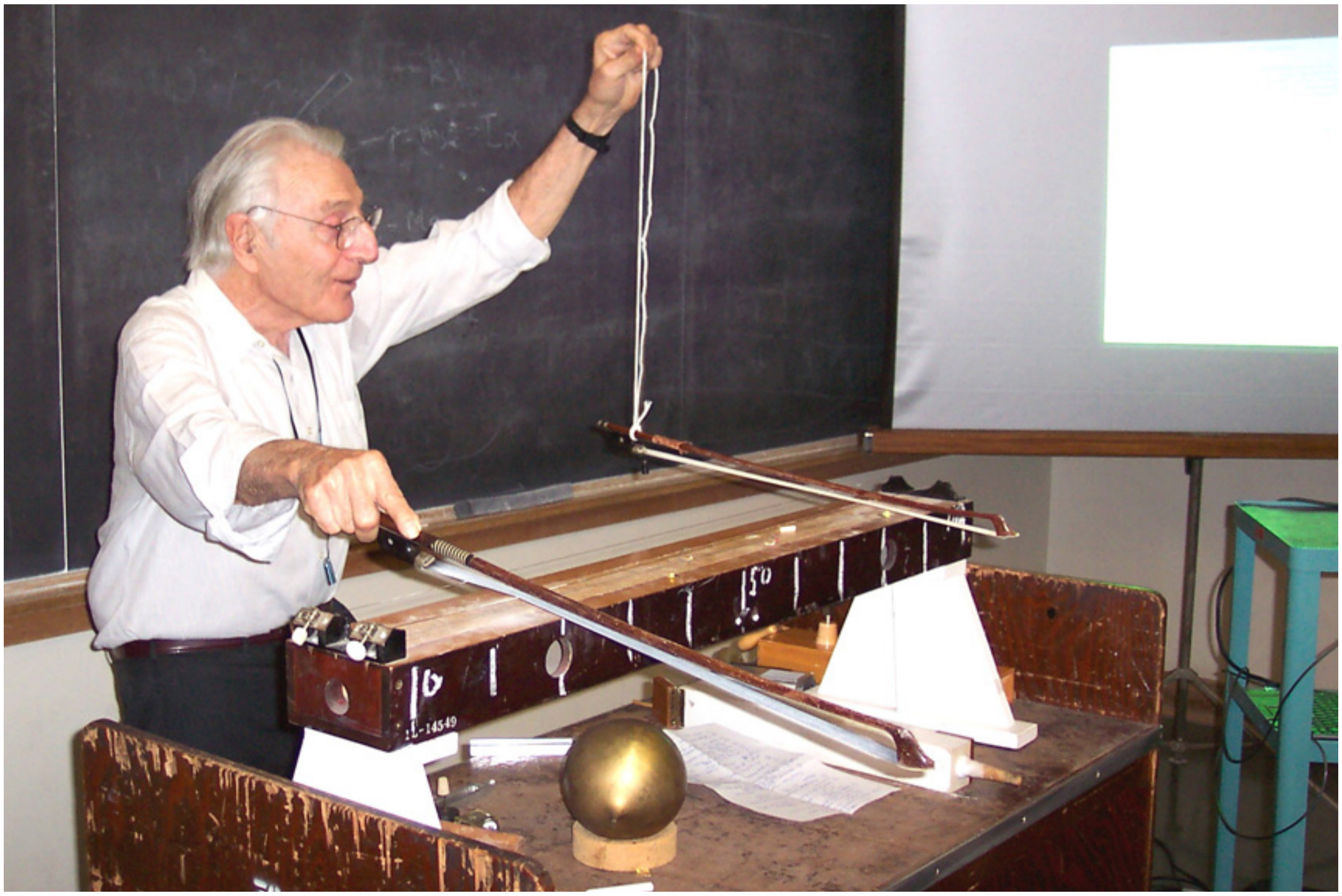

*Figure 5 Erwin delivering a lecture on Physics of Music (March 2004).*

When I came to the Department one day and proudly announced the birth of my daughter, Erwin said: "Congratulations, you are finally a productive scientist!"

And here are a couple of stories from other sources.

Famously, Erwin Hahn was quoted by a colleague as saying, "There is nothing that nuclear spins will not do for you, as long as you treat them as human beings."

At a social event in Asilomar where an NMR conference was running in parallel with a congress of Christian Scientists, Erwin was approached with a question: "Professor Hahn, are you a Christian Scientist?" "No, ma'am" was the answer, "I am a Jewish scientist."

## Concluding remarks

With his unique combination of exceptional innovation and physical intuition with a strong, jovial charisma and sense of humor, E. L. Hahn was a unique contributor to condensed matter physics and nonlinear optics. He will be missed, but his personality will be remembered by those who knew him and his contributions will continue to impact all of us.

## Acknowledgements

Some of the material in this article is adapted with permission from the International EPR(ESR) Society from EPR newsletter **21**(2), 6-8 (2011) (**www.epr-newsletter.ethz.ch**). We are indebted to Prof. J.-C. Diels for some of the original material, to the members of Prof. Hahn's family, and in particular to Katherine Hahn Halbheer and Elizabeth Hodgson, for their invaluable help in preparation of this article, and to Prof. L. Wald for sharing his memories of Prof. Hahn and for his important critical suggestions on the manuscript.

## List of selected influential publications of E. L. Hahn